\documentclass{article}

\usepackage{arxiv}

\usepackage[utf8]{inputenc} % allow utf-8 input
\usepackage[T1]{fontenc}    % use 8-bit T1 fonts
\usepackage{hyperref}       % hyperlinks
\usepackage{url}            % simple URL typesetting
\usepackage{booktabs}       % professional-quality tables
\usepackage{amsfonts}       % blackboard math symbols
\usepackage{nicefrac}       % compact symbols for 1/2, etc.
\usepackage{microtype}      % microtypography
\usepackage{lipsum}		% Can be removed after putting your text content
\usepackage{graphicx}
\usepackage{natbib}
\usepackage{doi}
\usepackage{graphicx}
\usepackage{caption}
\usepackage{float}
\usepackage{subcaption}
\usepackage{parskip}
\usepackage{titlesec}
\title{Project Proposal}

\author{{Alanna ~Hazlett} \\
	University of Virginia\\
        School of Data Science\\
        uwa6xv\\
	\texttt{uwa6xv@virginia.edu} \\
	\And
	{Naomi ~Ohashi} \\
	University of Virginia\\
        School of Data Science\\
        fju4ek\\
	\texttt{fju4ek@virginia.edu} \\
        \And
	{Timothy~Rodriguez} \\
	University of Virginia\\
        School of Data Science\\  
        tar3kh\\
	\texttt{tar3kh@virginia.edu} \\
        \And
	{Sodiq~Adewole} \\
	University of Virginia\\
        School of Data Science\\  
        soa2wg\\
	\texttt{soa2wg@virginia.edu} \\
}

\title{Chest Disease Detection in X-ray Images Using Deep Learning Classification Method}
\renewcommand{\undertitle}{Project Category: Image Classification with Convolutional Neural Networks}
\renewcommand{\shorttitle}{X-ray Classification Using Deep Learning}

\begin{document}
\maketitle

%\begin{abstract}
%	\lipsum[1]
%\end{abstract}

% keywords can be removed

%\section{Introduction}
%\lipsum[2]
%\lipsum[3]
% \section{Rivanna Repository}

% standard/sds\_managed\_sadewole/DS6050\_SP25/group3/

\section{Abstract}
In this work, we investigate the performance across multiple classification models to classify chest X-ray images into four categories of COVID-19, pneumonia, tuberculosis (TB), and normal cases. We leveraged transfer learning techniques with state-of-the-art pre-trained Convolutional Neural Networks (CNNs) models. We fine-tuned these pre-trained architectures on a labeled medical x-ray images. The initial results are promising with high accuracy and strong performance in key classification metrics such as precision, recall, and F1 score. We applied Gradient-weighted Class Activation Mapping (Grad-CAM) for model interpretability to provide visual explanations for classification decisions, improving trust and transparency in clinical applications.
\section{Motivation}
%\label{sec:headings}
%Detecting upper respiratory diseases using X-rays is a common practice. However, X-ray images can often be misleading. The application of Deep Learning Convolution Neural Network (CNN) techniques to detect and classify respiratory diseases from X-ray images supports clinicians in making decisions.
Detecting upper respiratory diseases using X-rays is common practice, especially when testing kits are either unavailable or results are delayed. However, interpreting X-ray images can be challenging, as it is often difficult to differentiate between various chest diseases.  Their interpretation can be subjective by clinicians and prone to errors, which underscores the need for accurate and automated diagnostic tools. 

The study aims to utilize Convolutional Neural Network (CNN) techniques and classify evidence of multiple respiratory diseases from X-ray images.  By applying neural networks, we seek to assist clinicians in quickly identifying respiratory diseases given limited time and lack of other testing resources. 

While this isn’t the first time Neural Networks have been used to detect respiratory diseases and some have achieved significant accuracy, we will attempt to expand the interpretability of the model’s results. We hope this increased understanding of how the model is making these classifications will spur more widespread acceptance and adoption of these types of models in practice.

\subsection{Research Question and Hypothesis}
To what extent can one improve the interpretability of highly accurate, high performing, and generalizable Convolutional Neural Networks (CNNs) for the purposes of identifying chest diseases in X-ray Images? 

%\subsection{Hypothesis}
We propose that through the implementation of explainability models such as Grad-CAM and others, one can develop a highly accurate, interpretable, lightweight model for the purposes of chest disease detection in X-ray images, which is deployable in low resource environments.

%\lipsum[4] See Section \ref{sec:headings}.

%\subsection{Headings: second level}
%\lipsum[5]

%\subsubsection{Headings: third level}
%\lipsum[6]

%\paragraph{Paragraph}
%\lipsum[7]

\section{Dataset}
\label{sec:headings}

\subsection{Description}
Originally, 46,737 chest X-ray images were collected from two Kaggle sources: “Chest X-ray (Pneumonia, COVID-19, TB, “CXR dataset”) and COVID-QU-Ex Dataset, “Rahman dataset.”  Additionally, 10,374 TB data samples were collected from the NIH NIAID TB Portal\cite{rosenthal_2017} to alleviate the imbalanced distribution. Access to NIH data required preapproval and a user agreement. This brings our total dataset size to 57,111 X-ray images across our training, validation, and test splits. The combined datasets are organized into three folders: train (41,219), validation (9,549), and test (6,343). Each folder contains four sub-folders for four classes: COVID-19 (8,542, 15.0\%), Normal (16,180, 28.3\%), Pneumonia (18,678, 32.7\%), and TB (13,711, 24.0\%).  We found collecting unique COVID-19 X-ray images that are publicly available to a be a significant challenge.  Some redundancies were removed from the final dataset.  Then, the dataset was normalized and resized to (224,224) or (299,299), depending on the transfer learning model used.   

\begin{table}[H]
    \centering
    \begin{tabular}{lcccccc}
        \hline
        Class & Train & Val & Test & Total \\
        \hline
        COVID-19 & 4,214 & 2,397 &  1,931 & 8,542 \\
        Normal & 11,448 & 3,517 &  1,215 & 16,180 \\
        Pneumonia & 14,589 & 2,264 &  1,825 & 18,678 \\
        Tuberculosis & 10,968 & 1,371 &  1,372 & 13,711 \\
        \hline
        Total & 41,219 & 9,549 & 6,343 & 57,111 \\
        \hline
    \end{tabular}
    \caption{Data Distribution}
    \label{Data Distribution}
\end{table}

%Here is an example usage of the two main commands (\verb+citet+ and \verb+citep+): Some people thought a thing \citep{kour2014real, hadash2018estimate} but other people thought something else \citep{kour2014fast}. Many people have speculated that if we knew exactly why \citet{kour2014fast} thought this\dots

\subsection{Data Source}
\begin{itemize}
    \item    
    \url{https://www.kaggle.com/datasets/jtiptj/chest-xray-pneumoniacovid19tuberculosis/data}\item     \url{https://www.kaggle.com/datasets/anasmohammedtahir/covidqu} \item
    \url{https://tbportals.niaid.nih.gov/}
\end{itemize}

\begin{figure}[htp]
    \centering
    \includegraphics[width=12cm]{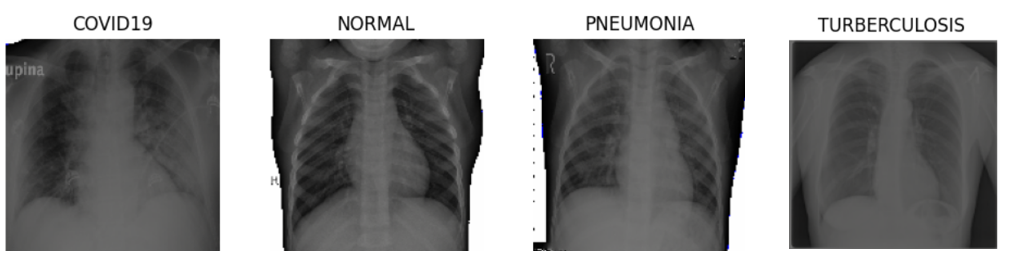}
    \caption{Chest X-ray images}
    \label{fig:CXR images}
\end{figure}

%See Figure \ref{fig:fig1}. Here is how you add footnotes. \footnote{Sample of the first footnote.}
%\lipsum[11]

%\begin{figure}
%	\centering
%	\fbox{\rule[-.5cm]{4cm}{4cm} \rule[-.5cm]{4cm}{0cm}}
%	\caption{Sample figure caption.}
%	\label{fig:fig1}
%\end{figure}

%\subsection{Tables}
%See awesome Table~\ref{tab:table}.

%The documentation for \verb+booktabs+ (`Publication quality tables in LaTeX') is available from:
%\begin{center}
%	\url{https://www.ctan.org/pkg/booktabs}
%\end{center}

%\begin{table}
%	\caption{Sample table title}
%	\centering
%%		\toprule
%		\multicolumn{2}{c}{Part}                   \\
%		\cmidrule(r){1-2}
%		Name     & Description     & Size ($\mu$m) \\
%		\midrule
%		Dendrite & Input terminal  & $\sim$100     \\
%		Soma     & Cell body       & up to $10^6$  \\
%		\bottomrule
%	\end{tabular}
%	\label{tab:table}
%\end{table}

%\subsection{Lists}
%\begin{itemize}
%	\item Lorem ipsum dolor sit amet
%	\item consectetur adipiscing elit.
%	\item Aliquam dignissim blandit est, in dictum tortor gravida eget. In ac rutrum magna.
%\end{itemize}

\section{Literature Survey}
% \label{sec:headings}
We conducted an extensive literature survey, selected the six most relevant articles, and reviewed each model's key approaches, methodology, and challenges.  All the models achieved over 90\% accuracy using convolutional neural network design compared to their benchmarks.  \cite{aravinda_2021} highlighted their model's explainability, which is essential for building trust in clinical settings, while \cite{thakur_2023} pointed out their benchmark needed better interpretability.  The common benchmarks included InceptionResNet, \cite{szegedy_2016}, ResNet \cite{he_2015}, and VGG19, \cite{simonyan_2014}.  These works leveraged the Keras package in order to generate their image classification models. Several of these works shared the common challenges of low-resolution images, overlapping of images, and/or image sensitivity. Further, several models were tested on relatively small datasets, raising questions about their generalizability.  
%\subsection{A demystifying convolutional neural networks using Grad-CAM for prediction of coronavirus disease (COVID-19) on X-ray images }
%\begin{center}
%	\url{https://pmc.ncbi.nlm.nih.gov/articles/PMC8137866/}
%\end{center}

The goal of \cite{aravinda_2021}'s study was to develop a quick and efficient method to pre-screen for COVID-19 using X-ray images, as an alternative to time-consuming PCR tests, \cite{aravinda_2021}. A convolutional neural network was built with 13 layers, including Conv2D, MaxPooling2D, and Dropout layers. It was trained on a dataset of 441 X-rays, augmented with various techniques like flipping and clipping. Grad-CAM was used to highlight the regions in the X-rays influencing the model's decisions, providing a method of explainability. Grad-CAM stores the prediction for the image and finds the gradients of the target class (COVID-19) score with respect to the feature maps of the last convolutional layer. The important neuron weights are determined by global average pooling the gradients. The activation map is multiplied to the pooled gradients. The class discriminative salience map is determined by the mean of the activation maps along the channels. ReLu is applied to the heatmap, so only features that have a positive influence on the output map are included. Then the intensity is normalized. The CNN achieved 96\% accuracy on the training set and 98\% on the validation set. While the model's strength lies in its explainability, its small dataset and architecture may limit its generalization. One area for study would be how generalizable is this model to other datasets and applications?  
%\subsection{COVID-Net: a tailored deep convolutional neural network design for detection of COVID-19 cases from chest X-ray images }
%\begin{center}
%	\url{https://www.nature.com/articles/s41598-020-76550-z}
%\end{center}

The COVID-Net model created by, \cite{wang_2020}, aims to quickly screen for COVID-19 using chest X-rays, which are widely available in health clinics. It classifies X-rays as normal, COVID-positive, or other illnesses using a deep convolutional neural network designed with both human and machine knowledge. The network follows the PEPX design pattern, reducing computational complexity through long-range connectivity. The model is pre-trained on ImageNet and fine-tuned on the COVIDx dataset of 13,975 X-rays. The data is preprocessed and augmented. GSInquire is used to provide explainability by highlighting key areas of the X-ray that influenced the model’s decision. While the model shows strengths in reducing computational complexity and offering transparency, it requires improvement in sensitivity and COVID prediction. It achieved 93.3\% test accuracy, 91\% sensitivity for COVID, and 98.9\% positive predictive value, outperforming VGG-19 and ResNet-50 in accuracy and predictive value but lagging in sensitivity. An additional avenue of exploration would be can one improve the sensitivity of this model while retaining the explainability achieved using GSInquire?

%\subsection{CDC\_Net: multi-classification convolutional neural network model for detection of COVID-19, pneumothorax, pneumonia, lung Cancer, and tuberculosis using chest X-rays }
%\begin{center}
%	\url{https://link.springer.com/article/10.1007/s11042-022-13843-7}
%\end{center}

\cite{malik_2023} introduced CDC\_Net, a deep convolutional neural network model designed for the multi-classification of chest diseases, including COVID-19, pneumothorax, pneumonia, lung cancer (LC), and tuberculosis (TB) using chest X-ray images.  The CDC\_Net model was compared with well-established pre-trained CNN models, such as Vgg-19, ResNet-50, and Inception v3 as a benchmark, to evaluate its performance in terms of classification accuracy, recall, precision, and F1-score.  The CDC\_Net model achieved an AUC of 0.9953 in the chest diseases, with an accuracy of 99.39\%, a recall of 98.13\%, and a precision of 99.42\%, outperformed other comparative models.  The authors addressed the challenges of low resolution and partial overlap in chest X-ray datasets for the CDC\_Net to capture different patterns of chest disease.  Discussion for other challenges such as computational cost and efficiency that the CDC\_Net model required could be critical for real-world deployment in resource-constraint environments. Could this model be improved via the use of skip connections or other methods to improve its computational efficiency? 

%\subsection{A Comprehensive Analysis of Deep Learning-Based Approaches for Prediction and Prognosis of Infectious Diseases }
%\begin{center}
%	\url{https://link.springer.com/article/10.1007/s11831-023-09952-7}
%\end{center}

\cite{thakur_2023} pursued the application of deep learning models for the prediction and prognosis of infectious diseases, focusing on the use of machine learning and deep learning techniques for infectious disease detection and classification.  A dataset of 29,252 images was compiled from various sources, encompassing eight classes: COVID-19, MERS, Pneumonia, SARS, Tuberculosis (TB), Viral Pneumonia, Lung Opacity, and Normal. The dataset was then subjected to exploratory data analysis, RGB histogram extraction, grayscale conversion, image augmentation, and contrast enhancement.  The authors evaluated pre-existing relevant deep learning models. Model performance was assessed using metrics such as accuracy, loss, F1 score, recall, precision, and RMSE.  The authors found that the InceptionResNetV2 model achieved the highest accuracy of 88\% on the testing dataset, with a loss of 0.399 and an RMSE of 0.63.  While the authors provided valuable insights into the use of deep learning for infectious disease detection, some limitations could have been addressed.  Complex deep learning models like InceptionResNetV2, are often considered "black boxes." The study did not address the interpretability of the models' predictions, which is crucial for clinical adoption. Could this model's interpretability be improved via implementation of an explainability model such as Grad-CAM? Further, could one improve this models accuracy through a reduction in the number of classes or additional training?  

%\subsection{COVID‑19 detection from chest X-ray images using transfer learning }
%\begin{center}
%	\url{https://www.nature.com/articles/s41598-024-61693-0#Sec3}
%\end{center}

\cite{elhouby_2024} investigated the use of pre-trained convolutional neural networks (CNNs) for the purpose of detection of SARS COV 2, also known as COVID-19. Houby implemented two CNN pre-trained models for binary classification (COVID/healthy) achieving experimental results with the best model of 95\% accuracy using enhanced full X-ray images. Both pre-trained models were pre-trained on the ImageNet dataset which contains more than a million images with over 1000 classes. The models were trained using a dataset containing 21,165 X-ray images, including “10,192 Normal cases, 3,616 positive COVID-19 cases, 1,345 Viral Pneumonia cases, and 6012 Lung Opacity” \cite{elhouby_2024}. Image enhancement techniques including Histogram Equalization, Contrast Limited Adaptive Histogram Equalization and Image Complement were applied to improve the image dataset for training and testing. The highest performing model was then applied to a four class, classification task (COVID-19, Normal, Viral Pneumonia-13 and Lung Opacity) and achieved accuracy of 93.5\%. The best performing model framework, VGG19, proved to be a promising model design for implementation across ever expanding datasets and the authors recommend its application to assist with real world diagnosis in the absence of readily available PCR testing. Could the interpretability of this model be improved by leveraging Grad-CAM or another explainability model?

%\subsection{Detection of COVID-19 Based on Chest X-rays Using Deep Learning }
%\begin{center}
%	\url{https://pmc.ncbi.nlm.nih.gov/articles/PMC8872326/}
%\end{center}

\cite{gouda_2022} utilized a pre-trained model, Resnet-50, modified with a skip connection layer allowing the layers' input signal to traverse the network by linking it to that layers output, for the purposes of 3 class, classification of X-ray images into COVID-19, Normal and Pneumonia. The dataset used for training comprised the COV-PEN dataset. Further, they augmented their dataset through the application of “masks such as rotation, reflection, shifting, and scaling” for every image of their dataset to enhance its size. Image enhancement techniques such as adaptive contrast enhancement were applied to the dataset. For training, they leveraged the “Adam and sigmoid optimizer with a learning rate strategy which decreased the learning rate when learning becomes stagnant”\cite{gouda_2022}.  The testing dataset was the combined benchmark of COVID-19 Image Data Collection (IDC) and CXR Images (Pneumonia) comprising approximately 558 X-ray images broken into the three classes mentioned. Their implementation yielded rather impressive results with their overall accuracy in this experiment of 99.63\%. This performance is beyond all previously published works at the time of its publication however, the rather small size of their testing dataset, around 558 X-ray images raises questions about it's generalizability.

 Several models we encountered in our research achieved extremely high levels of accuracy, but their generalizability is in question due to the small datasets they relied on. A few of the papers we reviewed focused on explanability, and while \cite{wang_2020} did use a large dataset, achieved high accuracy, and had explainability in mind, their work suffered from sensitivity issues compared to competing models. Thus, it stands to reason that an opportunity exists for our project to enhance the currently published literature to combine an explainable model with one that retains the accuracy, sensitivity, and generalizability in a relatively computationally light package to enable its application in limited resource settings.
\setlength{\parskip}{0pt}
\section{Method}
\subsection{Multiclass Convolutional Neural Networks}
Our task is inherently a classification task as we seek to identify which group our chest X-ray images into one of our four classes: COVID-19, normal, pneumonia, and tuberculosis. As such, we leverage supervised deep learning with convolutional neural networks (CNN) to achieve this classification task. CNNs have proven very adept at processing/classifying and identifying objects in images making them perfect for this particular task. This is due to the convolutional layers which employ kernels that traverse across the pixels of an image detecting certain patterns, for example edges, corners, or textures. In the case of classifying chest X-rays this ability is key as it allows the model to recognize patterns in the images that suggest one disease over another, learn those patterns, and then apply them to new data. 

In order to achieve high accuracy we leverage existing model architectures and weights by employing transfer learning. Transfer learning consists of leveraging the parameter weights and architecture of existing models to incorporate their knowledge into a new network. For our task, we leverage several existing model weights and architectures to achieve significantly higher accuracy than a standard fully trainable model would be able to achieve. The existing literature we reviewed made extensive use of transfer learning to achieve accuracy north of 90\%\ in nearly every instance. Thus, in order to match or exceed this state of the art accuracy we decided to implement transfer learning as well. Transfer learning further allows us to take advantage of the pre-trained model weights despite the fact that we have a comparably small dataset of X-ray images when we consider that most of the pre-trained models we employ were trained on ImageNet which contains more than 1.2 million images. Since our dataset consists of just 57,111 images across train, validation, and test splits, transfer learning provides a significant boost in performance compared to a fully bespoke model.

Finally, we take advantage of hyperparameter tuning to find the best setup for our models. Hyperparameter tuning allows us to identify the optimal setup for our model to ensure the highest performance. While this process could be performed manually, or we could just use commonly implemented values for all of our hyperparameters, implementing hyperparameter tuning allows us to procedurally determine the best settings for our models. 

\begin{figure}[H]
    \centering
    \includegraphics[width=12cm]{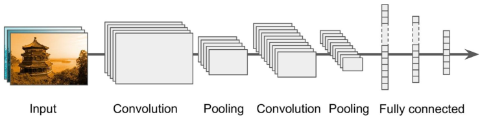}
    \caption{CNN Architecture}
    \label{CNN Architecture}
    Convolutional neural networks architecture with convolution, polling, and fully connected layer\cite{gron_2019}
\end{figure}

\subsection{Visualization Method}
% Introduction of Grad-CAM, the intended use
We applied Gradient-weighted Class Activation Map, known as Grad-CAM, for our visualization method to obtain a class activation heatmap for an image classification model. The Grad-CAM heatmap indicates warmer colors (red/yellow) highlight areas that a model predicts with the highest likelihood, while cooler colors (blue) indicate areas of less importance. Areas without the heatmap indicate that it is not part of the explanation for the classification. 

\begin{figure}[H]
    \centering
    \begin{minipage}[b]{0.49\textwidth} 
        \centering
        \includegraphics[height=5cm]{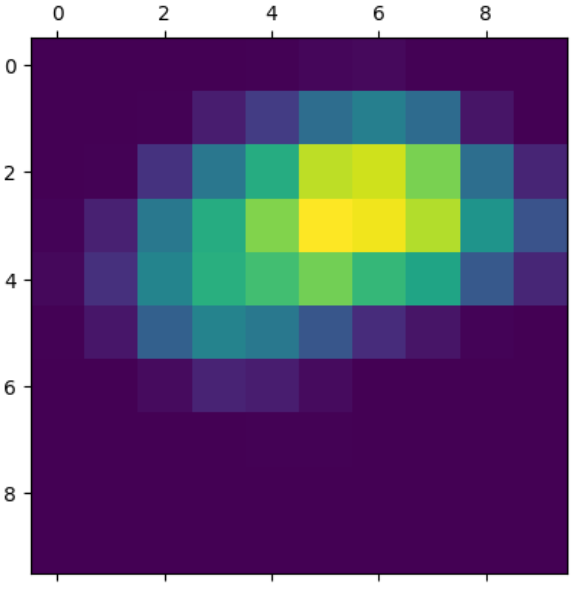}
        \caption*{(a) Heatmap}
    \end{minipage}
    \hfill
    \begin{minipage}[b]{0.49\textwidth}
        \centering
        \includegraphics[height=5cm]{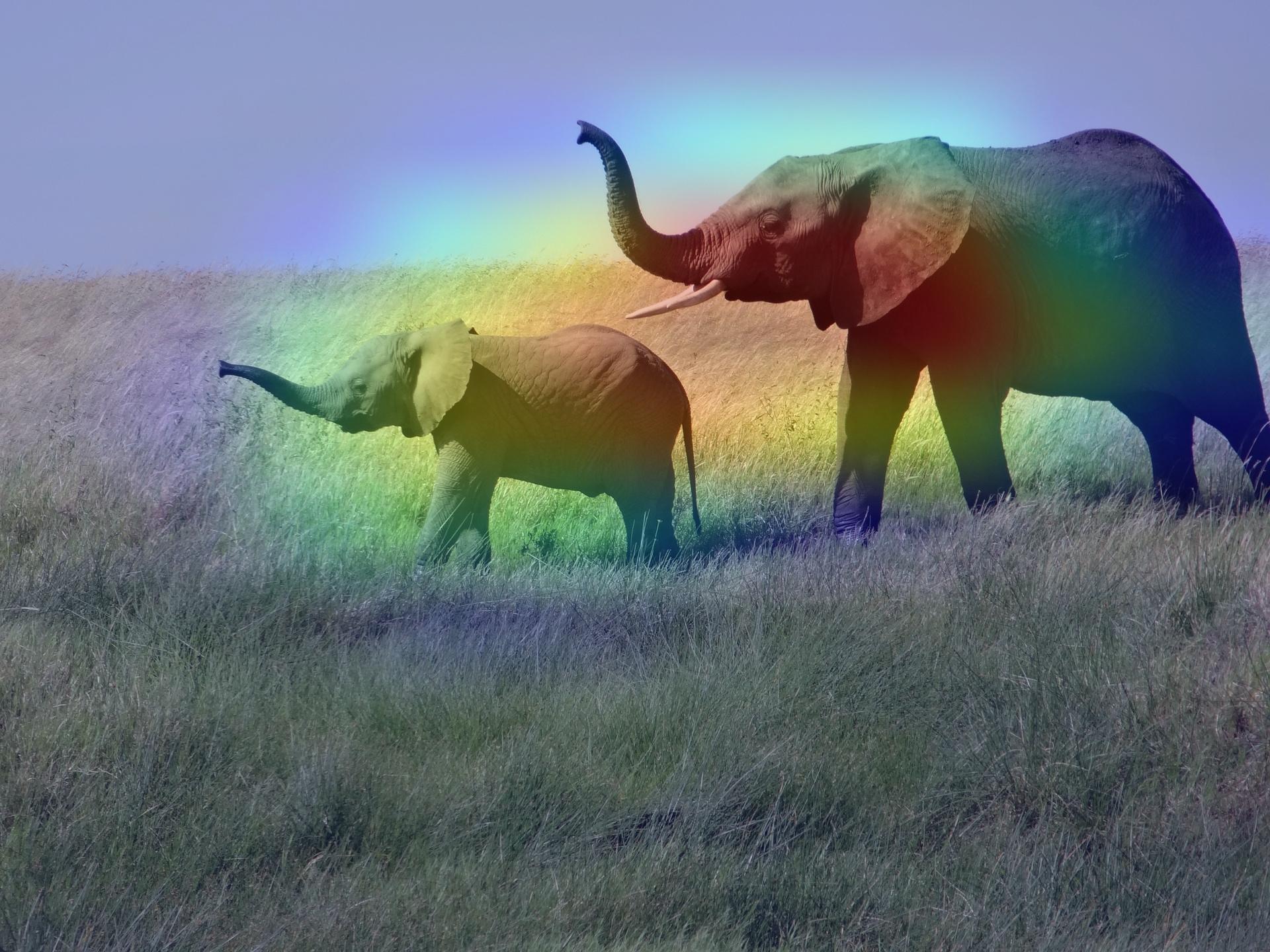}
        \caption*{(b) Grad-CAM Heatmap Elephant}
    \end{minipage}
    
    \vspace{0.5em}
    
    \small Source: \url{https://keras.io/examples/vision/grad_cam/}
    
    \caption{Grad-CAM Heatmap Application}
    \label{fig:figure4}
\end{figure}

% \begin{figure}[H]
%     \centering
%     \begin{subfigure}[b]
%         %\centering
%         \includegraphics[width=\textwidth]{grad_cam_10_0.jpg}
%         \caption{Grad-CAM Heatmap Application}
%         \label{Grad-CAM Heatmap Application}
%     \end{subfigure}
%     \hfill
%     \begin{subfigure}[b]
%         %\centering
%         \includegraphics[width=12cm]{grad_cam_heatmap_elelephant.png}
%         \caption{Grad-CAM Heatmap Application}
%         \label{Grad-CAM Heatmap Application}
%     \end{subfigure}
%     \vspace{0.5cm}

%     \small source by:\url{https://keras.io/examples/vision/grad_cam/}
% \end{figure}

\setlength{\parskip}{0pt}
\section{Results}

Our experiments resulted in the achievement of state of the art accuracy, AUC,  F1-score, recall, and precision using our bespoke dataset of 57,111 X-ray images across our four classes of COVID-19, Normal, Pneumonia, and TB. As seen in Figure 4 we achieve more than 96\% accuracy across all of our base model architectures. Since this is the largest dataset employed for this specific four class, classification task, we can be relatively confident these models will retain their scalability may have reasonably strong generalization.

\begin{table}[H]
    \centering
    \begin{tabular}{lcccccc}
        \hline
        Pre-Trained Model Architecture & Accuracy & Loss & AUC & F1-Score & Recall & Precision \\
        \hline
        ResNet50 & 97.99\% & 0.0718 &  99.73\% & 97.98\% & 97.95\% & 98.02\%\\
        VGG16 & 96.44\% & 0.1131 &  99.57\% & 96.33\% & 96.40\% & 96.47\%\\
        Xception & 97.12\% & 0.1198 &  94.60\% & 97.02\% & 99.51\% & 69.80\%\\
        EfficientNetV2B0 & 98.24\% & 0.0962 &  99.73\% & 98.11\% & 98.24\% & 98.24\%\\
        \hline
    \end{tabular}
    \caption{Test Dataset Results - All Models}
    \label{Test Results - All Models}
\end{table}

\setlength{\parskip}{0pt}
\subsection{Training Results}

 We trained each transfer learning model for 10 epochs utilizing early stopping where applicable. All four models showed continued improvement on the training datasets with respect to accuracy and loss with minimal evidence of overfitting. Performance on the validation dataset was generally stable during training for all models besides the Xception model which displayed instability early on.

%Test metrics table here
% \begin{figure}[H]
%     \centering
%     \begin{minipage}[b]{0.49\textwidth} 
%         \centering
%         \includegraphics[height=5cm]{grad_cam_heatmap_elelephant.png}
%         \caption*{(a) Heatmap}
%     \end{minipage}
%     \hfill
%     \begin{minipage}[b]{0.49\textwidth}
%         \centering
%         \includegraphics[height=5cm]{grad_cam_10_0.jpg}
%         \caption*{(b) Grad-CAM Heatmap Elephant}
%     \end{minipage}
    
%     \vspace{0.5em}
    
%     \small Source: \url{https://keras.io/examples/vision/grad_cam/}
    
%     \caption{Grad-CAM Heatmap Application}
%     \label{fig:figure4}
% \end{figure}

\begin{figure}[H]
    \centering
    \begin{minipage}[b]{0.49\textwidth}
        \includegraphics[width=8cm]{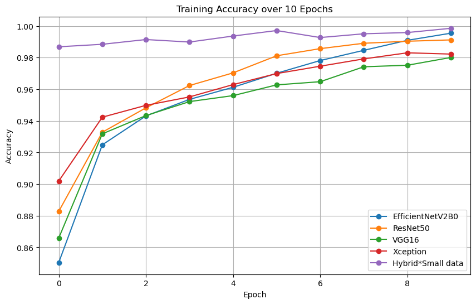}
        \caption{Training Accuracy over 10 Epochs}
        \label{Training Accuracy over 10 Epochs}      
    \end{minipage}
    \hfill
    \begin{minipage}[b]{0.49\textwidth}
        \includegraphics[width=8cm]{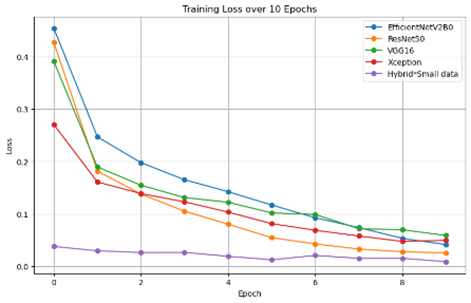}
        \caption{Training Loss over 10 Epochs}
        \label{Training Loss over 10 Epochs}
    \end{minipage}
    
     \label{fig:figure4}
\end{figure}
A hybrid model (ResNet50 and VGG16) was reproduced from a Kaggle submission using the same chest X-ray images with the initial dataset (7,135 images), shown in purple in Figures 4 and 5.  We attempted to train the hybrid model with our large dataset,  however, the model estimated over 6 hours to train one epoch using an A-100 GPU. The heavy compute required for the hybrid architecture represents a major challenge for the scalability and practicality of using this model on our research project's dataset. Therefore, we concluded this model is not practical to apply to our use case. We included this model in our charts to display a baseline performance comparison despite the smaller dataset used.

\subsection{Best Model}
To select the best model among our transfer learning models, we use accuracy and AUC from the validation set.  While ResNet50 and EfficientNetV2B0 have a similar pattern in both accuracy and AUC, ResNet50 has steady improvements and generally outperforms the other models. Further on our test dataset Resnet50 returned strong performance in all metrics resulting in the second highest accuracy behind only EfficientNetV2B0.  Based on the results on both the validation and test datasets, we concluded that the ResNet50 model is the strongest model among the models we tested.

\begin{figure}[H]
    \centering
    \begin{minipage}[b]{0.49\textwidth}
        \includegraphics[width=8cm]{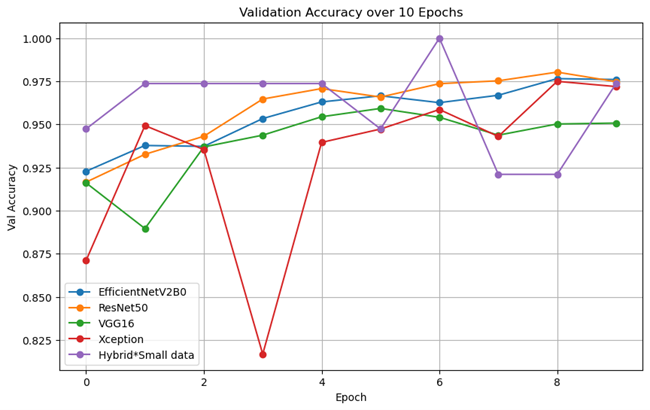}
        \caption{Validation Accuracy over 10 Epochs}
        \label{Validation Accuracy over 10 Epochs}      
    \end{minipage}
    \hfill
    \begin{minipage}[b]{0.49\textwidth}
        \includegraphics[width=8cm]{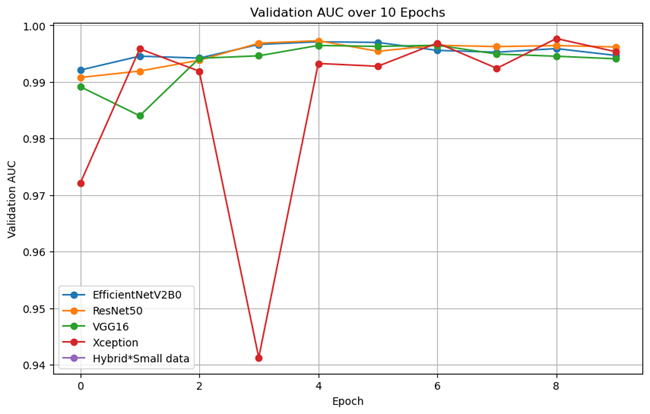}
        \caption{Validation AUC over 10 Epochs}
        \label{Validation AUC over 10 Epochs}
    \end{minipage}
    
     \label{fig:figure4}
\end{figure}

% Conclusion of the overall performance
\subsection{SME vs Grad-CAM (VGG16)}

We collaborated with a medical student at the University of California Irvine, Timothy McMullen PhD, for a blind review and classification of several of the images in our test dataset to compare his annotations and classifications with those of our various models. Dr. McMullen identified several areas in the TB image seen on the left, circled in green and red, in Figure 8 which indicated the presence of disease. Several of these same areas were identified by the VGG16 model's Grad-CAM heatmap. While the Grad-CAM heatmap and Dr. McMullen's annotations do not align perfectly, upon review of the Grad-CAM image, Dr. McMullen expressed confidence that in this instance the model had reasonably identified many of key areas for diagnosis of this image. It should be noted that X-ray images are seldom used in isolation when making a diagnosis and a list of patient symptoms and medical history is typically available to the attending physician. Dr. McMullen suggested that a classification model like the ones we produced for this project could be used in conjunction with language models that ingest patient data including symptom lists in order to more accurately aid in diagnosis. This could be a potential area for future study. 

\begin{figure}[H]
    \centering
    \begin{subfigure}[b]{0.35\textwidth}
        \includegraphics[width=5cm, height=5cm]{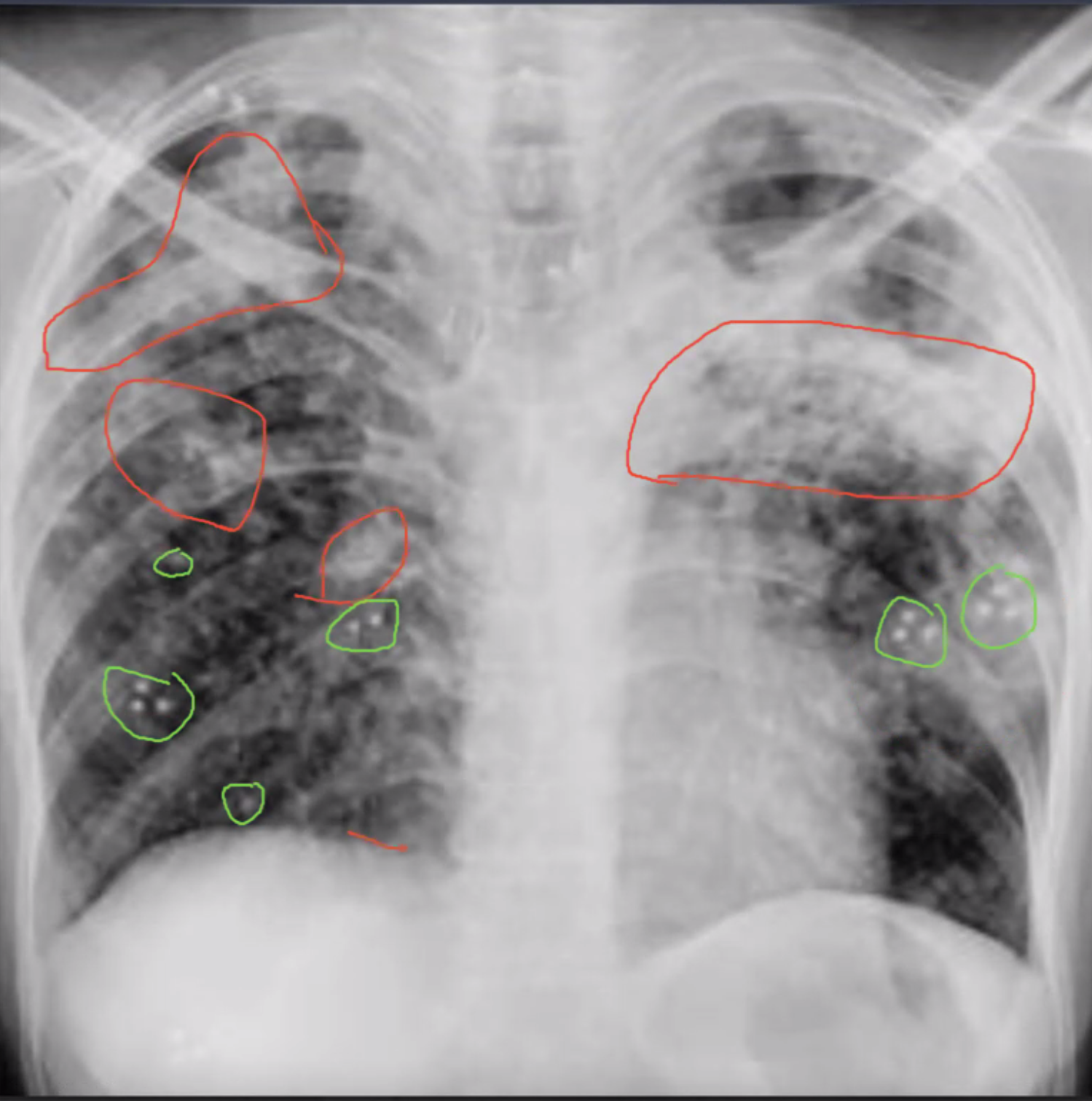}
    \end{subfigure}
    \begin{subfigure}[b]{0.35\textwidth}
        \includegraphics[width=5cm, height=5cm]{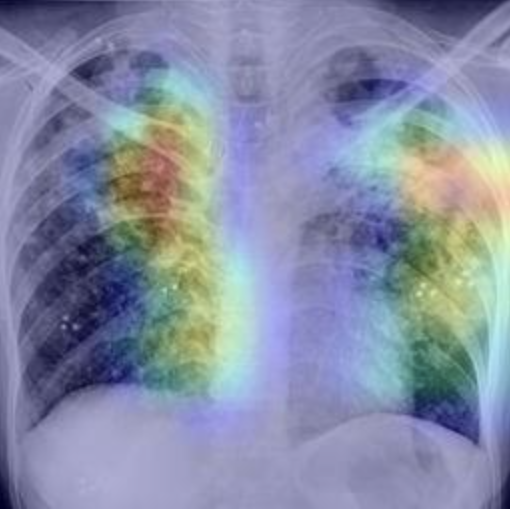}
    \end{subfigure}
    \caption{SME Evaluation of TB X-ray (left) vs Grad-CAM with VGG16 (right)}
\end{figure}

\subsection{ResNet50}
\subsubsection{Description of Model:} 

ResNet50 has 50 layers which starts with a convolution layer and pooling and then is followed by 4 major stages, cov2, conv3, conv4, and conv5. Each of these stages has multiple residual blocks. These then contain batch normalization, ReLu activation function, and skip connections. ResNet50 ends with a max pooling layer. We built upon this architecture by using a flatten layer and two dense layers. The last 10 layers of ResNet50 were trainable. This resulted in a final parameter count of 24,708,612 parameters of which  4,537,476 were trained leaving 20,171,136 parameters not trained. 

%\begin{table}[H]
%    \centering
%    \begin{tabular}{lcccccc}
%        \hline
%        Size & Top-1 Accuracy & Top-5 Accuracy & Parameters & Depth %& Time per inference step (GPU) \\
%        \hline
%        98 MB & 74.9\% & 92.1\% &  25.6M & 107 & 4.6ms\\
%        \hline
%    \end{tabular}
%    \caption{ResNet50}
%    \label{ResNet50}
%\end{table}

%\subsubsection{Mentioned by these papers:} 
%\begin{itemize}
%\item Demystifying convolutional neural networks using grad-CAM for prediction of %coronavirus disease (COVID-19) on X-ray images 
%\item Comprehensive analysis of deep learning-based approaches for prediction and %prognosis of infectious diseases
%\item COVID-net: a tailored deep convolutional neural network
%design for detection of COVID-19 cases from chest X-ray images
%\item CDC\_Net: multi-classification convolutional neural network model for detection %of COVID-19, pneumothorax, pneumonia, lung Cancer, and tuberculosis using chest X-rays
%\item COVID-19 detection from chest X-ray images using transfer learning
%\item Detection of COVID-19 Based on Chest X-rays Using Deep Learning
%\end{itemize}
%\subsubsection{Experiments:} 

We utilized transfer learning of ResNet50 by adding a couple of dense layers. We used hyperparameter tuning to determine number of neurons for these layers, activation function for these layers, and learning rate. We experimented with the number of layers of ResNet50 that were trainable. Grad-CAM was utilized to provide explainability to the model's classification and prediction. 
 
\subsubsection{Predictions and Grad-CAM:}

\begin{figure}[H]
    \centering
    \includegraphics[width=10cm, height=7cm]{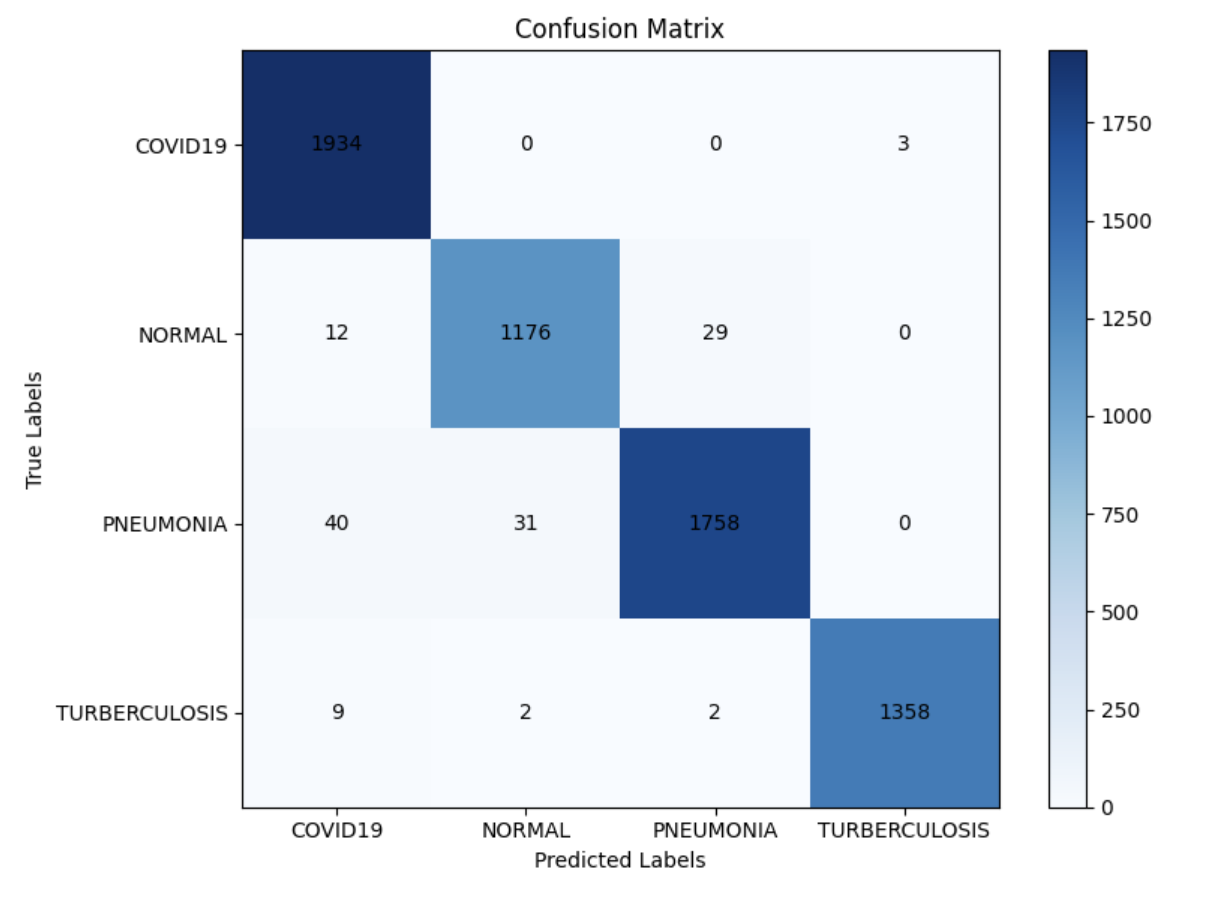}%{cov_matrix_resnet50.png}
    \caption{Confusion Matrix, ResNet50}
    \label{Training and Validation Results, ResNet50}
\end{figure}

\begin{figure}[H]
    \centering
    \begin{subfigure}[b]{0.48\textwidth}
        \includegraphics[width=\linewidth, height=4cm]{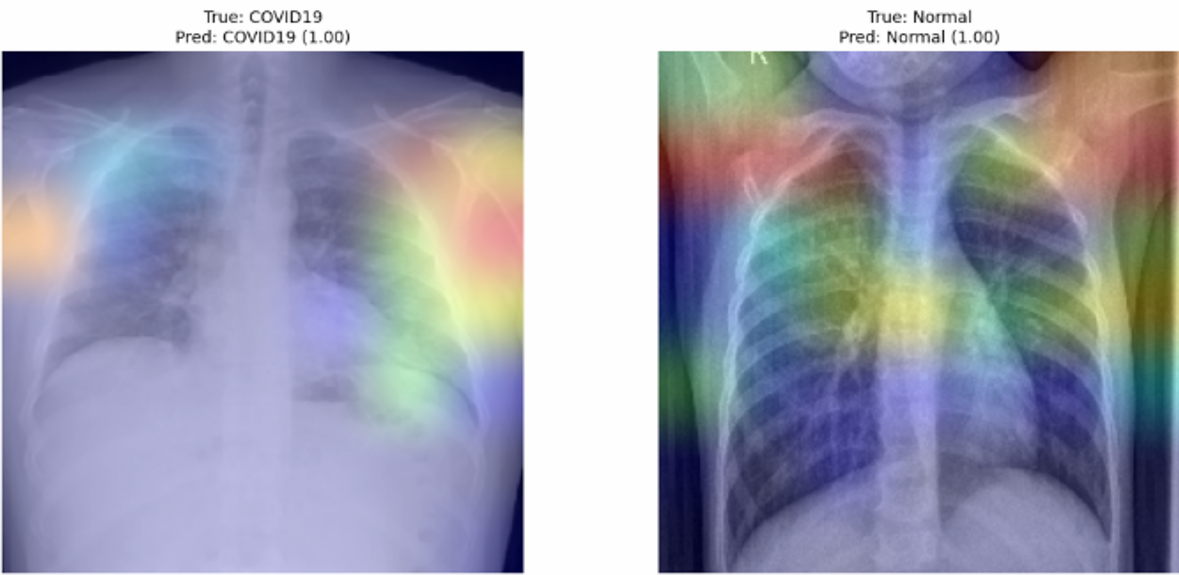}
    \end{subfigure}
    \begin{subfigure}[b]{0.48\textwidth}
        \includegraphics[width=\linewidth, height=4cm]{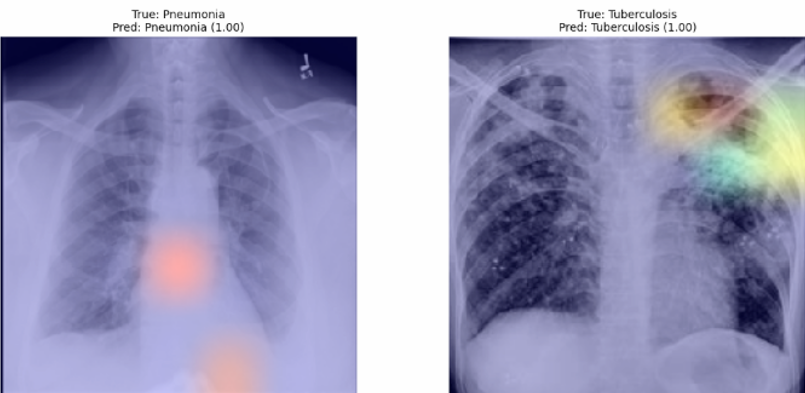}
    \end{subfigure}
    \caption{First 4 test images with prediction and Grad-CAM overlay, ResNet50}
\end{figure}

%\begin{figure}[H]
%    \centering
%    \includegraphics[width=10cm, height=10cm]{ResNet50_First_4.png}
%    \caption{First 4 test images with prediction and Grad-CAM overlay, ResNet50}
%    \label{First 4 images (test set) with prediction and GradCAM overlay, Resnet50}
%\end{figure}

%\subsubsection{Training, Validation, and Test Results:} 

%\begin{table}[H]
 %   \centering
 %   \begin{tabular}{lcccccc}
 %       \hline
 %       Metric & loss & accuracy & auc & f1\_score & Recall & %Precision \\
 %       \hline
 %       Training & 0.0546 & 0.9803 & 0.9990 & 0.9835 & 0.9802 & 0.9807 %\\
 %       Validation & 0.3281 & 0.9076 & 0.9820 & 0.8988 & 0.9033 & %0.9113 \\
 %       Test & 0.3327 & 0.9038 & 0.9818 & 0.9104 & 0.9009 & 0.9073 \\
 %       \hline
 %   \end{tabular}
 %   \caption{Training, Validation, and Test dataset results, ResNet50}
 %   \label{Training, Validation, and Test dataset results}
%\end{table}

%\begin{figure}[H]
%    \centering
%    \includegraphics[width=12cm]{resenet50_train_val.png}
%    \caption{Training and Validation Results, ResNet50}
%    \label{Training and Validation Results, Xception}
%\end{figure}

\subsection{VGG16}
\subsubsection{Description of Model:}

VGG16 has 16 layers that has five blocks, which are comprised of two or three convolutional layers followed by a max pooling layer. The convolutional layers use a 3x3 kernel and have ReLU activation functions. After all of the blocks the model concludes with a max pooling layer. We built upon this architecture by using a flatten layer and two dense layers. The last 10 layers of VGG16 were trainable. This resulted in a final parameter count of 15,076,612 parameters of which 13,341,124 were trained leaving 1,735,488 parameters not trained.
%\begin{table}[H]
%    \centering
%    \begin{tabular}{lcccccc}
%        \hline
%        Size & Top-1 Accuracy & Top-5 Accuracy & Parameters & Depth & %Time per inference step (GPU) \\
%        \hline
%        528MB & 71.3\% & 90.1\% &  138.4M & 16 & 4.2ms\\
%        \hline
%    \end{tabular}
%    \caption{VGG16}
%    \label{VGG16}
%\end{table}

%\subsubsection{Mentioned by these papers:} 
%\begin{itemize}
%\item Demystifying convolutional neural networks using grad-CAM for prediction of %coronavirus disease (COVID-19) on X-ray images
%\item Deep Residual Learning for Image Recognition 
%\item CDC\_Net: multi-classification convolutional neural network model for detection %of COVID-19, pneumothorax, pneumonia, lung Cancer, and tuberculosis using chest X-rays
%\item COVID-19 detection from chest X-ray images using transfer learning
%\item Detection of COVID-19 Based on Chest X-rays Using Deep Learning
%\end{itemize}
%\subsubsection{Experiments:} 

We utilized transfer learning of VGG16 by adding a couple of dense layers. We used hyperparameter tuning to determine number of neurons for these layers, activation function for these layers, and learning rate. We experimented with the number of layers of VGG16 that were trainable. Grad-CAM was utilized to provide explainability to the model's classification and prediction. 

\subsubsection{Predictions and Grad-CAM:} 

%Utilizing 8 images from our test set, one from each class, the model successfully classified all 8. 

%\vspace{-2em}
\begin{figure}[H]
    \centering
    \includegraphics[width=10cm, height=7cm]{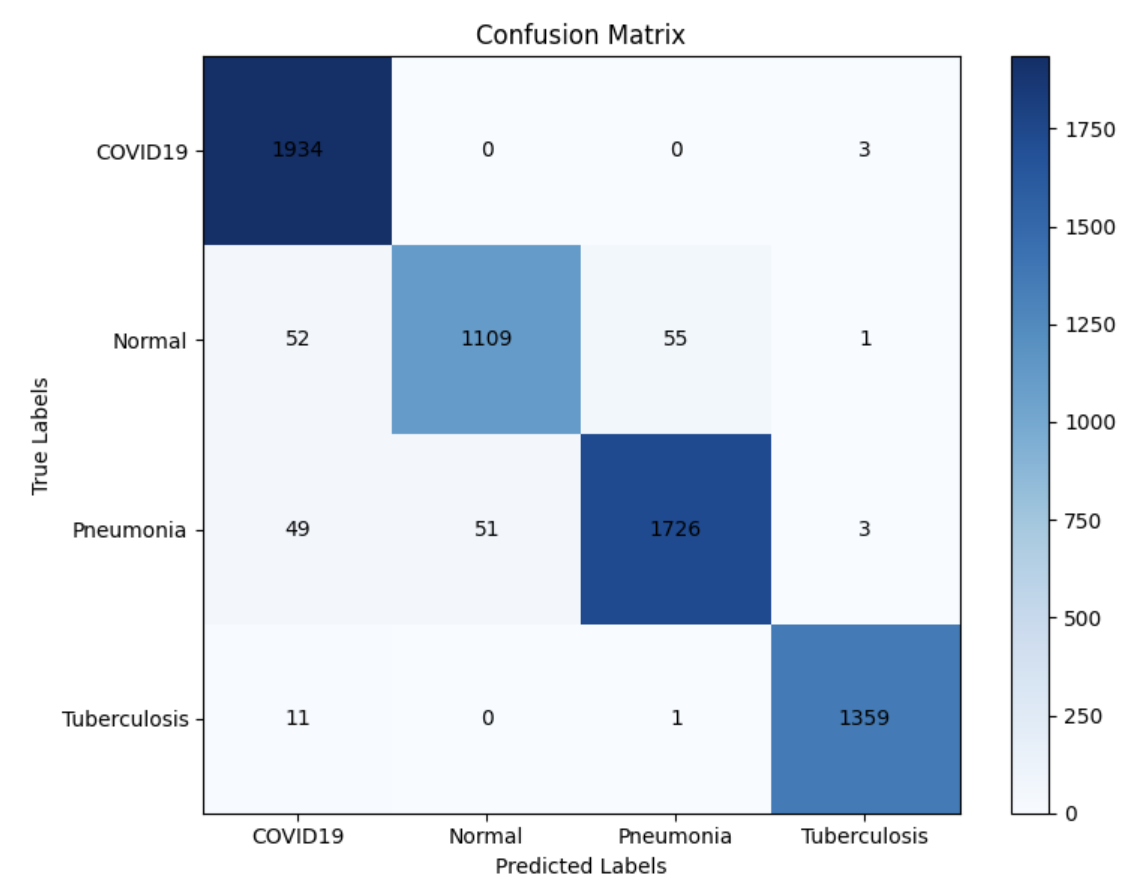}
    \caption{Confusion Matrix, VGG16}
    \label{Confusion Matrix VGG16}
\end{figure}
%\vspace{-1em}

%\begin{figure}[H]
%    \centering
%    \begin{subfigure}[b]{0.25\textwidth}
%    \caption*{True: COVID-19 \\ Pred: COVID-19 (1.00)}
%        \includegraphics[width=4cm, height=4cm]{VGG16_First_COVID.png}
%    \end{subfigure}
%    \begin{subfigure}[b]{0.25\textwidth}
%    \caption*{True: Normal \\ Pred: Normal (1.00)}
%        \includegraphics[width=4cm, height=4cm]{VGG16_First_Normal.png}
%    \end{subfigure}
%    
%    \vspace{0.1cm}
%    
%    
%    \begin{subfigure}[b]{0.25\textwidth}
%   \caption*{True: Pneumonia \\ Pred: Pneumonia (1.00)}
%       \includegraphics[width=4cm, height=4cm]{VGG16_First_Pneumonia.png}
%    \end{subfigure}
%    \begin{subfigure}[b]{0.25\textwidth}
%    \caption*{True: Tuberculosis \\ Pred: Tuberculosis (1.00)}
%       \includegraphics[width=4cm, height=4cm]{VGG16_First_Tuberculosis.png}
%    \end{subfigure}
%    
%    \caption{First 4 test images with prediction and Grad-CAM overlay, VGG16}
%    \label{fig:first4images}
%\end{figure}

\begin{figure}[H]
    \centering
    \begin{subfigure}[b]{0.24\textwidth}
        \caption*{True: COVID-19 \\ Pred: COVID-19 (1.00)}
        \includegraphics[width=\linewidth, height=4cm]{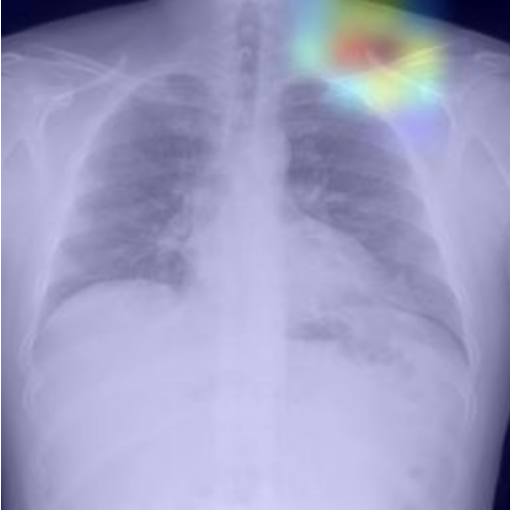}
    \end{subfigure}
    \begin{subfigure}[b]{0.24\textwidth}
        \caption*{True: Normal \\ Pred: Normal (1.00)}
        \includegraphics[width=\linewidth, height=4cm]{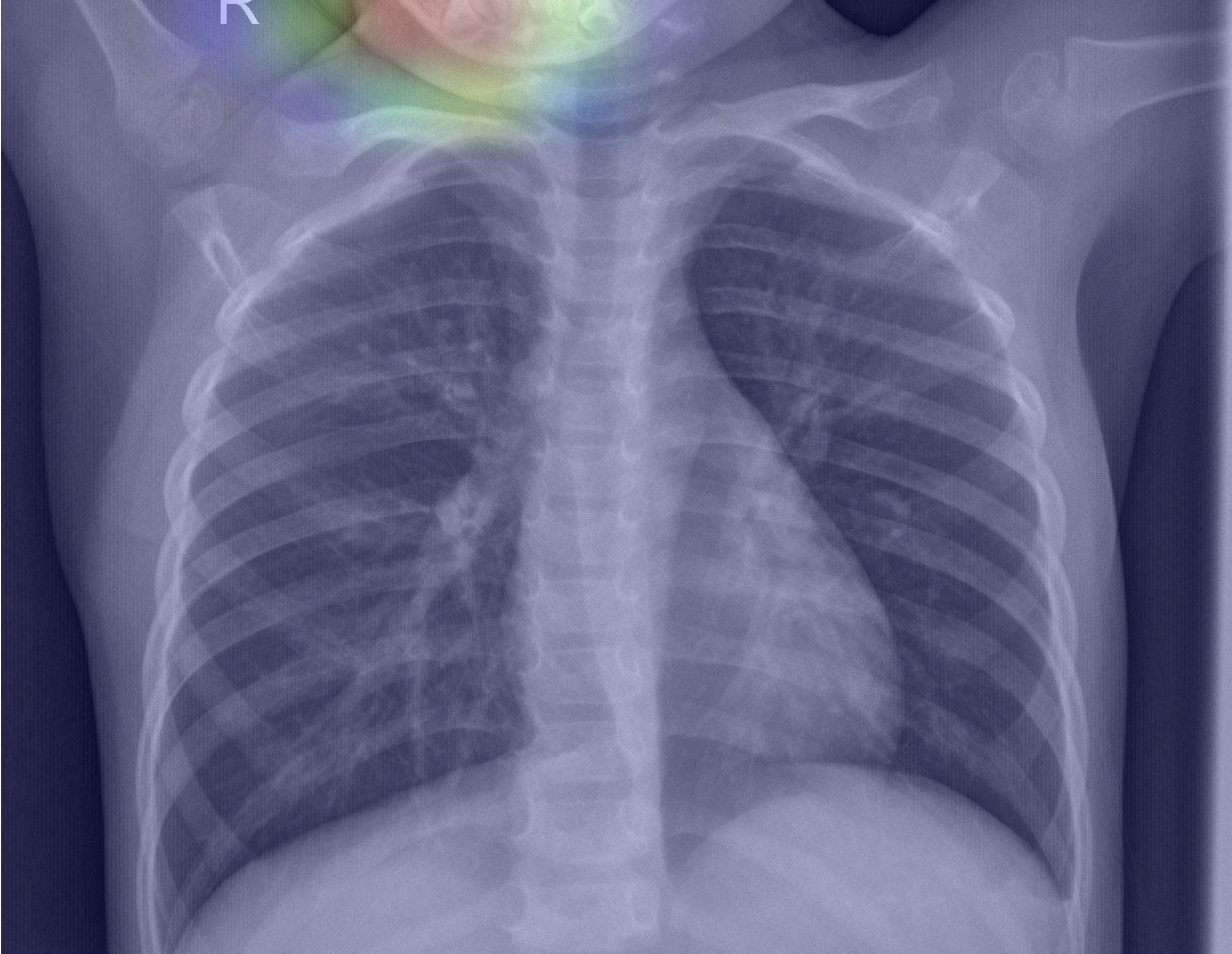}
    \end{subfigure}
    \begin{subfigure}[b]{0.24\textwidth}
        \caption*{True: Pneumonia \\ Pred: Pneumonia (1.00)}
        \includegraphics[width=\linewidth, height=4cm]{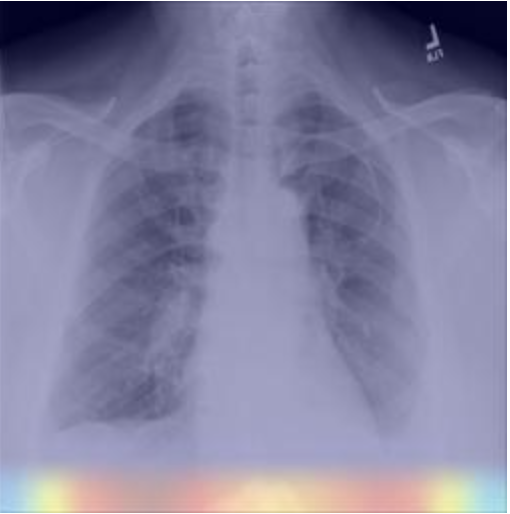}
    \end{subfigure}
    \begin{subfigure}[b]{0.24\textwidth}
        \caption*{True: Tuberculosis \\ Pred: Tuberculosis (1.00)}
        \includegraphics[width=\linewidth, height=4cm]{VGG16_First_Tuberculosis.png}
    \end{subfigure}
    
    \caption{First 4 test images with prediction and Grad-CAM overlay, VGG16}
    \label{fig:first4images}
\end{figure}

%\subsubsection{Training, Validation, and Test Results:} 

%\begin{table}[H]
%    \centering
%    \begin{tabular}{lcccccc}
%        \hline
%        Metric & loss & accuracy & auc & f1\_score & Recall & Precision \\
%        \hline
%        Training & 0.0561 & 0.9797 & 0.9992 & 0.9830 & 0.9794 & 0.9802 \\
%        Validation & 0.1717 & 0.9483 & 0.9930 & 0.9536 & 0.9481 & 0.9497 \\
%        Test & 0.1340 &	0.9521 & 0.9957 & 0.9526 & 0.9516 & 0.9534 \\
%        \hline
%    \end{tabular}
%    \caption{Training, Validation, and Test dataset results, VGG16}
%    \label{Training, Validation, and Test dataset results}
%\end{table}

%\begin{figure}[H]
%    \centering
%    \includegraphics[width=12cm]{VGG16_Model5Plots.png}
%    \caption{Training and Validation Results, VGG16}
%    \label{Training and Validation Results, VGG16}
%\end{figure}

\titlespacing*{\section}{0pt}{1ex minus .2ex}{1ex minus .2ex}

\subsection{Xception}
\setlength{\parskip}{0pt}
\subsubsection{Description of Model:} 

The Xception model employs a depthwise separable convolutional neural network architecture. Xception consists of an entry flow mainly comprised several regular convolution layers, and some depthwise layers, then a middle flow which consists of eight repetitions of depthwise separable convolution layers with ReLU activation and finally an exit flow consisting of increasing sized separable convolution, global average pooling and finally a fully connected layer for output. Our research slightly extended this architecture by adding a flattening layer and another dense fully connected layer with 64 nodes. This resulted in a final parameter count of 20,992,876 parameters of which we trained 20,938,348, leaving only 54,528 parameters non-trainable.
\setlength{\parskip}{0pt}
%\begin{table}[H]
%    \centering
%    \begin{tabular}{lcccccc}
%        \hline
%        Size & Top-1 Accuracy & Top-5 Accuracy & Parameters & Depth & Time per inference step (GPU) \\
%        \hline
%        88 MB & 79.0\% & 94.5\% &  22.9M & 81 & 8.1ms\\
%        \hline
%   \end{tabular}
%    \caption{Xception}
%    \label{Xception}
%\end{table}

%\subsubsection{Mentioned by these papers:} 
%\begin{itemize}
%\item CDC\_Net: multi-classification convolutional neural network model for detection of COVID-19, %pneumothorax, pneumonia, lung Cancer, and tuberculosis using chest X-rays
%\item COVID‑19 detection from chest X-ray images using transfer learning
%\end{itemize}
\titlespacing*{\section}{0pt}{1ex minus .2ex}{1ex minus .2ex}
\subsubsection{Experiments:} 
\setlength{\parskip}{0pt}

We utilized transfer learning of Xception by adding a couple of dense layers. We conducted several experiments varying the neuron layers including hyperparameter tuning to determine the number of neurons for these layers, activation function for these layers, and learning rate. However, none of the hyperparameter tuning results were able to match the performance initial experiment hyperparameter setup. We experimented with the number of layers of Xception that were trainable. Grad-CAM was utilized to provide explainability to the model's classification and prediction.

\subsubsection{Predictions and Grad-CAM:} 

Our confusion matrix details the very high accuracy this model achieved on the test dataset. The largest group of misclassifications were in the Pneumonia and Normal sections with 121 images predicted to be Normal when they were actually Pneumonia. 

\begin{figure}[H]
    \centering
    \includegraphics[width=10cm, height=7cm]{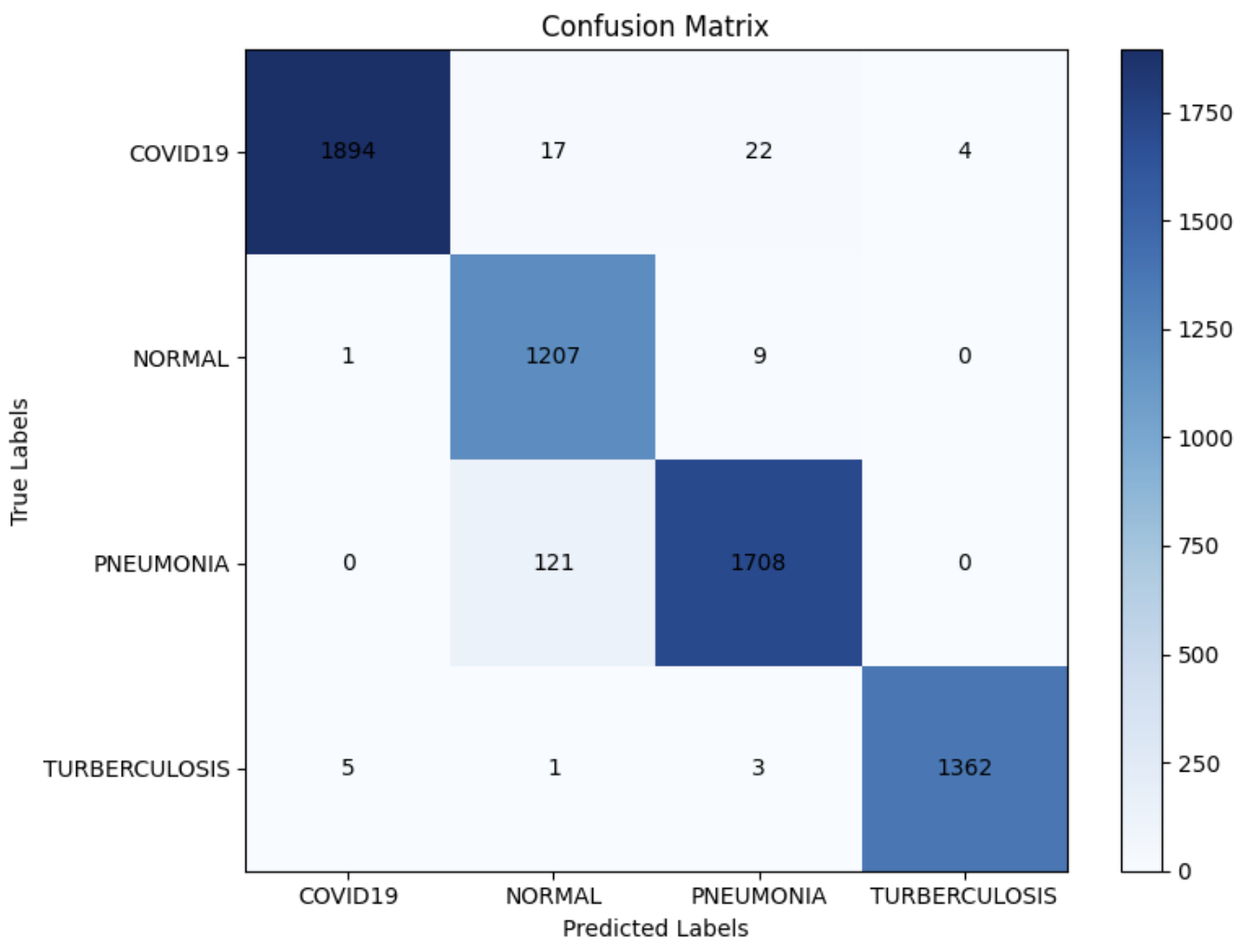}
    \caption{Confusion Matrix, Xception}
    \label{Confusion Matrix (Test set), Xception}
\end{figure}

%\begin{figure}[H]
%    \centering
%    \includegraphics[height=9cm]{xception First 4 Images gradcam.png}
%    \caption{First 4 test images with prediction and Grad-CAM overlay, Xception}
%    \label{First 4 images (test set) with prediction and GradCAM overlay, Xception}
%\end{figure}

\begin{figure}[H]
    \centering
    \begin{subfigure}[b]{0.48\textwidth}
        \includegraphics[width=\linewidth, height=4cm]{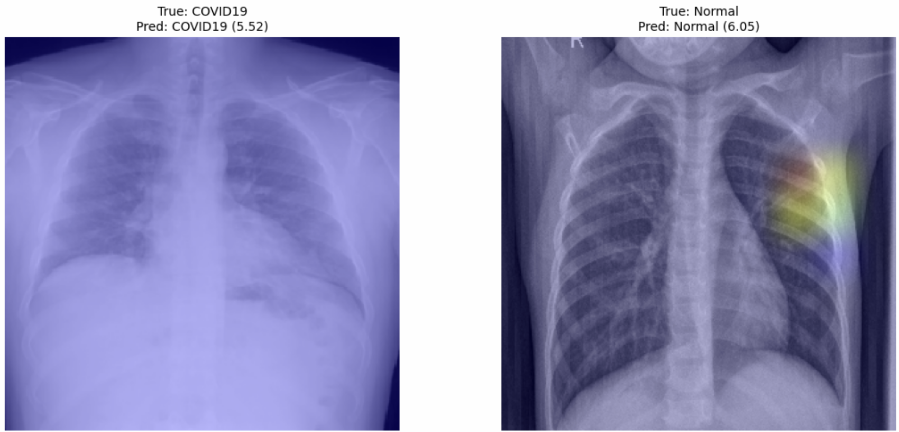}
    \end{subfigure}
    \begin{subfigure}[b]{0.48\textwidth}
        \includegraphics[width=\linewidth, height=4cm]{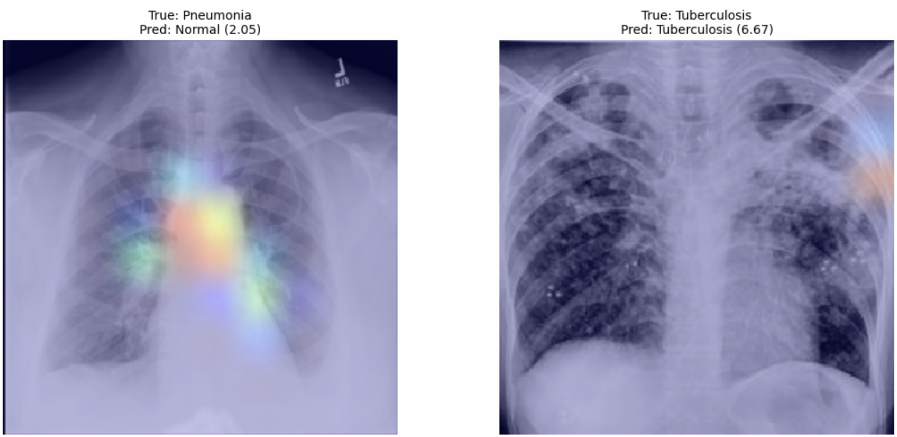}
    \end{subfigure}
    \caption{First 4 test images with prediction and Grad-CAM overlay, Xception}
\end{figure}

%\subsubsection{Training and Validation Results:} 

%\begin{figure}[htp]
%    \centering
%    \includegraphics[width=12cm]{xception_train_val.png}
%    \caption{Training and Validation Results, Xception}
%    \label{Training and Validation Results, Xception}
%\end{figure}

\subsection{EfficientNetV2B0}

\subsubsection{Description of Model:} 
EfficientNetV2 are a family of convolutional neural networks classification models, which is more robust and efficient than its previous version, EfficientNetV1. The models use neural architecture search (NAS) to optimize model size and faster training \cite{tan_2021}. While scaling up, the architecture achieved both training speed and good accuracy by adjusting regularization, such as dropout and data augmentation.  It uses ImageNet weight loading, input shape, pooling, output classes, activation function, and returns a Keras model. Notablly, EfficientNetV2 models expect their inputs to be float tensors of pixels with values in the [0, 255] range.

% \begin{table}[H]
%     \centering
%     \begin{tabular}{lcccccc}
%         \hline
%         Size & Top-1 Accuracy & Top-5 Accuracy & Parameters & Depth & Time per inference step (GPU) \\
%         \hline
%         29 MB & 77.1\% & 93.3\% &  5.3M & 132 & 4.9ms\\
%         \hline
%     \end{tabular}
%     \caption{EfficientNetB0}
%     \label{EfficientNetB0}
% \end{table}

%\subsubsection{Mentioned by these papers:} 

%\begin{itemize}
%\item EfficientNet: Rethinking Model Scaling for Convolutional Neural Networks \
%\item EfficientNetV2: Smaller Models and Faster Training
%\end{itemize}

\subsubsection{Experiments:}
This architecture uses EfficientNetV2b0 as a frozen base model without the top classification layer and uses ImageNet pretrained weights to transfer learn.  The input layer receives 224 x 224 x 3 images, which are preprocessed to match EfficientNetV2 requirements.  The inputs are passed through the frozen base model to extract Global Average Pooling 2D, reducing them to a vector.  To enhance and prevent overfitting, a Dropout layer (0.4 rate) is followed by a dense layer with 225 neurons (ReLU activation) and L2 regularization.   Then, a dense output layer with 4 neurons and softmax activation produces the class probabilities for a 4-class classification problem.  During model training, Early Stopping and Reduce On Plateau functions are applied.  During the last run, Early Stopping went through the entire 10 epochs.  The learning rate at 1e-4 remained until the 9th epoch, then switched to 5e-5 at the 10th epoch.  This resulted in a final parameter count of 6,249,300 parameters of which we trained 3,189,236, leaving 3,060,064 parameters non-trainable.

The Grad-CAM class activation visualization offers ImageNet utilities, "decode predictions" to interpret predictions from the models trained on ImageNet, such as ResNet50 and VGG16.  However, the decode prediction function does not work with the EfficientNetV2 models.  To workaround this issue, a new definition for "decode prediction" was created. 

\subsubsection{Predictions and Grad-CAM:} 
The confusion matrix indicates the model performs well by classifying true positives correctly.  Normal is misclassified as pneumonia (47), and pneumonia is misclassified as normal (45).  This is a common misclassification between those two conditions due to the visual similarities in chest X-rays.  COVID-19 has minimal confusion, showing strong model differentiation.

\begin{figure}[htp]
    \centering
    \includegraphics[width=10cm, height=7cm]{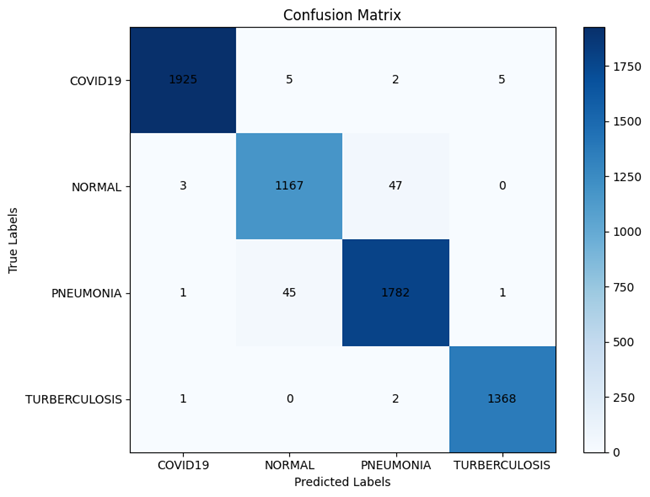}
    \caption{Confusion Matrix EfficientNetB0}
    \label{Confusion Matrix EfficientNetB0}
\end{figure}

\begin{figure}[H]
    \centering
    \begin{subfigure}[b]{0.48\textwidth}
        \includegraphics[width=\linewidth, height=4cm]{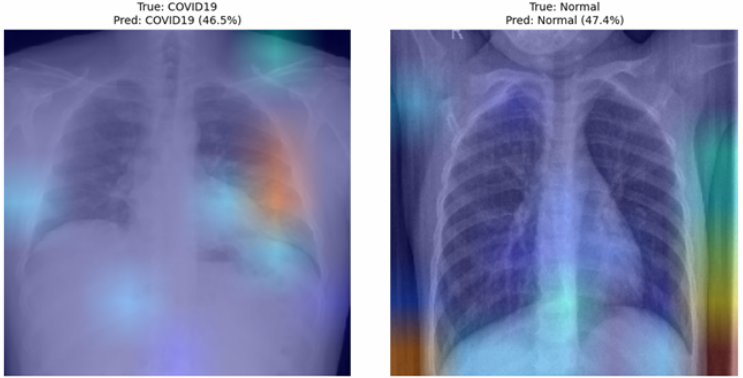}
    \end{subfigure}
    \begin{subfigure}[b]{0.48\textwidth}
        \includegraphics[width=\linewidth, height=4cm]{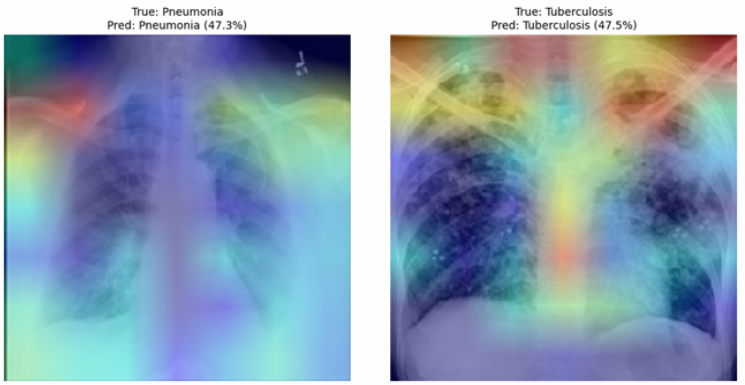}
    \end{subfigure}
    \caption{First 4 test images with prediction and Grad-CAM overlay, EfficientNetB0}
\end{figure}

\subsection{Overall Results of Grad-CAM:} 

Across all of the models we tested, Grad-CAM provided insights into where in each X-ray image the model identified to be the key areas driving the particular classification prediction that model selected. In some cases, as we discussed in section 7.3, Grad-CAM revealed that the model identified areas broadly similar to those identified by a medical student with experience identifying chest diseases in X-ray images. However, we also noted several instances in which Grad-CAM identified areas outside the lungs as the primary locations driving predictions. These included several images which highlighted the edges of each image, or the corner of the X-ray where a black box was located which covered up personally identifiable information (PII). Others, identified areas of tissue on the arms which for chest disease identification is irrelevant from a medical perspective. These instances suggest that there are features of the dataset and of individual classes that are being identified by the model for classification that are diagnostically irrelevant. This finding could suggest that while our models did perform well on our hold out test dataset, we may find limitations in generalization due to these models learning features of our dataset that will not be present in X-rays fed into the model in the future. 

We encountered numerous issues with the implementation of the Grad-CAM package which resulted in differing heatmaps displaying depending on whether or not a single image or a group of images was displayed at once. Future enhancements to this research should include a focus on improving the stability and consistency of the Grad-CAM results and displays. 

\section{Conclusion}

Through our investigation of the implementation of CNNs for the purpose of classifying chest diseases in X-ray images, we discovered that it is possible to improve the explainability of high performing generalizable models. We found that in several instances Grad-CAM, when applied to our model set, is able to highlight areas of the lungs that those with medical training also identified as indicators of particular diseases or indications of a healthy patient. Further, while all of our models were able to achieve state of the art accuracy on our large, bespoke, dataset, we found that our ResNet50 based model to be our best model across validations and test sets showcasing stable training and strong generalization. 

However, both the explainability and generalizability of our models have limitations due to certain idiosyncrasies in the datasets we leveraged. Grad-CAM revealed that for some X-ray images, features outside the lungs including the masking of personally identifiable information in certain classes in the dataset served as an indicator for the model as to which class the image belonged. While our dataset, which was larger, better balanced and more diverse than other datasets we encountered in comparable literature, certain features of particular classes (such as many images of the normal class being children), resulted in the model using features outside the lungs to get the correct classification.

In summary, while we did find some success improving both the generalizability and explainability of our classification models, there remains work to be done before these models are ready to be deployed in a diagnostic environment.

\section{Areas for Future Study}

In the course of this investigation we noted several areas for future study that we encourage others to pursue. We decided initially to enlarge our dataset in order to improve generalization and ensure balanced classes as many of the datasets we encountered were quite small and very imbalanced. However, we discovered that many of our images would benefit from the implementation of image pre-processing to remove markers that our models are learning to identify a particular class rather than using the chest X-ray image itself. This behavior became apparent in some of the Grad-CAM images which highlighted areas outside the lungs as key for classification. We also suggest that future work programmatically evaluate Grad-CAM heatmaps across many more examples in the test dataset to understand misclassifications and identify potential biases in the dataset to address through either augmentation or further image pre-processing. We also noted that several of the classes of our dataset were beset by varying image sizes and qualities which degrade the performance of the model. We recommend that future work seek to standardize and improve the quality of images, and standardize the distribution of patient age across the dataset classes to prevent the model learning signal from poor image quality or the age of the patient. This re-balancing of the dataset could enhance the generalizability of these models. 

Finally, our SME Dr. McMullen, highlighted that for diagnostic medicine, patient profiles and symptoms are available for doctors to make informed decisions. As such one additional area to explore would be the integration of these classification model predictions into a large language model (LLM) which is trained on and provided with patient data to further assist doctors with quick and accurate diagnoses.

\section{Acknowledgment}
Some data were obtained from the TB Portals (https://tbportals.niaid.nih.gov), which is an open-access TB data resource supported by the National Institute of Allergy and Infectious Diseases (NIAID) Office of Cyber Infrastructure and Computational Biology (OCICB) in Bethesda, MD. These data were collected and submitted by members of the TB Portals Consortium \url{https://tbportals.niaid.nih.gov/Partners}. Investigators and other data contributors that originally submitted the data to the TB Portals did not participate in the design or analysis of this study. "TB Portals Published CXRs" \url{https://datasharing.tbportals.niaid.nih.gov/#/data-collection-overview} was used in the analysis described here.

We would like to thank Professor Sodiq Opeyemi Adewole, PHD, our mentor, Ahson Saiyed, and our SME, Timothy McMullen, PHD, for their invaluable guidance throughout this research project.

\bibliographystyle{unsrtnat}
\bibliography{references}

%\bibliography{references2}  %%% Uncomment this line and comment out the ``thebibliography'' section below to use the external .bib file (using bibtex) .

%\begin{thebibliography}{1}




%\end{thebibliography}

\end{document}